\documentclass[amsmath,amssymb,twocolumn]{revtex4}
\usepackage{graphicx}
\usepackage{amscd,amsmath,amsthm,amsfonts,amssymb}
\def\[{\begin{equation}}
\def\]{\end{equation}}

\begin{document}
\title{Partial-rogue waves that come from nowhere but leave with a trace in the
Sasa-Satsuma equation}
\author{
Bo Yang$^{1}$, Jianke Yang$^{2}$}
\address{$^{1}$ School of Mathematics and Statistics, Ningbo University, Ningbo 315211, China\\
$^{2}$ Department of Mathematics and Statistics, University of Vermont, Burlington, VT 05401, U.S.A}
\begin{abstract}
Partial-rogue waves, i.e., waves that ``come from nowhere but leave with a trace", are analytically predicted and numerically confirmed in the Sasa-Satsuma equation. We show that, among a class of rational solutions in this equation that can be expressed through determinants of 3-reduced Schur polynomials, partial-rogue waves would arise if these rational solutions are of certain orders, where the associated generalized Okamoto polynomials have real but not imaginary roots, or imaginary but not real roots. We further show that, at large negative time, these partial-rogue waves approach the constant-amplitude background, but at large positive time, they split into several fundamental rational solitons, whose numbers are determined by the number of real or imaginary roots in the underlying generalized Okamoto polynomial. Our asymptotic predictions are compared to true solutions, and excellent agreement is observed.
\end{abstract}
\maketitle

\section{Introduction}
Rogue waves has been a subject of intensive theoretical and experimental studies in mathematical and physical communities in the past decade. Hundreds of papers and several books have been published on it, and more are still coming. Rogue waves are often defined as ``waves that come from nowhere and leave without a trace" \cite{Akhmediev2009}. For example, they can be localized wave excitations that arise from the constant-amplitude background, reach higher amplitude, and then retreat back to the same background, as time progresses. Almost all rogue waves that have been theoretically derived or experimentally observed belong to this category (see \cite{Peregrine,AAS2009,DGKM2010,Fiber1,Tank1}, among many others).

However, there exists another type of waves that ``come from nowhere but leave with a trace". Specifically, these waves also arise from the constant-amplitude background (thus ``come from nowhere"), stay localized, and reach higher amplitude. Afterwards, instead of retreating back to the same constant background with no trace, they evolve into localized waves on the constant background that persist at large time, thus leaving a trace. The first report of such peculiar waves seems to be in \cite{OhtaYangDSII} for the Davey-Stewartson-II equation, where a two-dimensional localized wave arose from the constant background and then split into two localized lumps at large time (see Fig.~4 of that paper). Later, a similar but one-dimensional solution was reported in \cite{Zhao16} for the Sasa-Satsuma equation. These peculiar waves resemble rogue waves in the first half of evolution, but contrast them in the second half of evolution. Due to these peculiar behaviors, let us call them partial-rogue waves. Note that although two examples of partial-rogue waves can be seen in \cite{OhtaYangDSII,Zhao16}, there was no explanation for their appearance, as if they were pure accidents. It was also unclear whether additional types of partial-rogue waves could be found in those two systems.

In this paper, we predict partial-rogue waves in the Sasa-Satsuma equation through large-time asymptotic analysis on its rational solutions. We show that, among a class of rational solutions in this equation that can be expressed through determinants of 3-reduced Schur polynomials, partial-rogue waves arise if and only if these rational solutions are of certain orders, where the associated generalized Okamoto polynomials have real but not imaginary roots, or imaginary but not real roots. We further show that, at large negative time, these partial-rogue waves approach the constant-amplitude background, but at large positive time, they split into several fundamental rational solitons, whose numbers are determined by the number of real or imaginary roots in the underlying generalized Okamoto polynomial. Our asymptotic predictions are compared to true solutions, and excellent agreement is observed.

\section{Preliminaries}
The Sasa-Satsuma equation was proposed as a higher-order nonlinear Sch\"odinger equation for optical pulses that includes some additional physical effects such as third-order dispersion and self-steepening \cite{SS1991,Kodama_Hasegawa1987}. Through a variable transformation, this equation can be written as
\[ \label{SS}
u_t=u_{xxx}+ 6|u|^2u_x + 3u(|u|^2)_x.
\]
Sasa and Satsuma \cite{SS1991} showed that this equation is integrable.

\subsection{A class of rational solutions} \label{sec_rational}
Soliton solutions on the zero background in this equation were derived by Sasa and Satsuma in their original paper \cite{SS1991}. Later, rational solutions on a constant background, including rogue waves, were also derived \cite{Zhao16,LingSS2016,AkhmedievSS12,Chen13,Zhao14,MuQin16,He19,MuQin20,Feng22,Feng22b}. The solutions that will be the starting point of this paper are a certain class of rational solutions which, in the language of Darboux transformation, are associated with a scattering matrix admitting a triple eigenvalue. Such solutions have been studied in \cite{LingSS2016,Zhao16} by Darboux transformation. However, their solutions are not general nor explicit for our purpose. For this reason, we will first present general and explicit expressions for this class of rational solutions through Schur polynomials.

Before presenting these solutions, we need to specify the nonzero background. Through variable scalings, we can normalize the background amplitude to be unity. Then, this background can be written as
\[ \label{bc}
u_{bg}(x, t)= e^{{\rm{i}}[\alpha(x + 6 t)-\alpha^3 t]},
\]
where $\alpha$ is a free wavenumber parameter, which cannot be removed since the Sasa-Satsuma equation (\ref{SS}) is not Galilean-invariant. But $\alpha$ can be restricted to be positive, since the Sasa-Satsuma equation is invariant under the axes reflection of $(x, t)\to (-x, -t)$, and negative-$\alpha$ solution can be related to positive-$\alpha$ solution through this axes reflection.

To present these explicit rational solutions, we also need to introduce elementary Schur polynomials. These polynomials $S_j(\mbox{\boldmath $x$})$ with $ \emph{\textbf{x}}=\left( x_{1}, x_{2}, \ldots \right)$ are defined by the generating function
\begin{equation}\label{Elemgenefunc}
\sum_{j=0}^{\infty}S_j(\mbox{\boldmath $x$}) \epsilon^j
=\exp\left(\sum_{j=1}^{\infty}x_j \epsilon^j\right).
\end{equation}
In addition, we define $S_{j} (\emph{\textbf{x}})=0$ when $j<0$.

Our expressions for general rational solutions corresponding to a triple eigenvalue in the scattering matrix of Darboux transformation are given by the following theorem.

\begin{quote}
\textbf{Theorem 1} \hspace{0.05cm} When $\alpha=1/2$, the Sasa-Satsuma equation (\ref{SS}) admits bounded $(N_1, N_2)$-th order rational solutions
\[
u_{N_1, N_2}(x,t)= \frac{g_{N_1, N_2}}{f_{N_1, N_2}}e^{{\rm{i}}[\alpha(x + 6 t)-\alpha^3 t]}, \label{Schpolysolu1}
\]
where $N_1$ and $N_2$ are arbitrary non-negative integers,
\[ \label{SchpolysolufN}
f_{N_1, N_2}=\sigma_{0,0}, \quad g_{N_1, N_2}=\sigma_{1,0},
\]
\[ \label{sigmaTheorem3}
\sigma_{k,l} = \det \left( \begin{array}{cc}
                           \sigma_{k,l}^{\left[1,1\right]} & \sigma_{k,l}^{\left[1,2\right]} \\
                           \sigma_{k,l}^{\left[2,1\right]} & \sigma_{k,l}^{\left[2,2\right]}
                         \end{array}
\right),
\]
\[\label{Blockmatrix}
\sigma^{[I, J]}_{k,l}=
\left(
\phi_{3i-I, \, 3j-J}^{(k,l \hspace{0.04cm} I, J)}
\right)_{1\leq i \leq N_{I}, \, 1\leq j \leq N_{J}},
\]
matrix elements in $\sigma^{[I, J]}_{k,l}$ are defined by
\begin{eqnarray} \label{Schmatrimnij9a}
&& \hspace{-1cm} \phi_{i,j}^{(k,l, I, J)}=\sum_{\nu=0}^{\min(i,j)} \left( \frac{p_{1}^2 }{4p_0^2}  \right)^{\nu}  \times  \nonumber \\
&& \hspace{-0.4cm} S_{i-\nu}(\textbf{\emph{x}}_I^{+}(k,l) +\nu \textbf{\emph{s}})  \hspace{0.06cm} S_{j-\nu}(\textbf{\emph{x}}_J^{-}(k,l) + \nu \textbf{\emph{s}}),
\end{eqnarray}
vectors $\textbf{\emph{x}}^{\pm}_{I}(k,l)=( x_{1,I}^{\pm}, x_{2,I}^{\pm},\cdots)$ are given by
\begin{eqnarray}
&&\hspace{-1.3cm} x_{r,I}^{+}(k,l)= p_{r} (x+6t) + \beta_{r} t + k\theta_{r}+ l \theta_{r}^*+ a_{r,I}, \label{defxrp2} \\
&&\hspace{-1.3cm} x_{r,J}^{-}(k,l)=   p_{r}  (x+6t)  + \beta_{r} t - k\theta_{r}^*- l \theta_{r} +a_{r,J},  \label{defxrm2}
\end{eqnarray}
$\beta_{r}$ and $\theta_{r}$ are coefficients from expansions
\begin{eqnarray}
&& \hspace{-1.1cm} p^3(\kappa) = \sum_{r=0}^{\infty} \beta_{r}\kappa^{r}, \hspace{0.09cm}
\ln \left[ \frac{ p \left( \kappa \right)+\textrm{i}\alpha}{p_{0}+\textrm{i}\alpha}\right] =\sum_{r=1}^{\infty} \theta_{r}\kappa^{r}, \label{schucoeflambda1}
\end{eqnarray}
the function $p\left( \kappa \right)$ with expansion $p\left( \kappa \right)=\sum_{r=0}^{\infty}p_{r} \kappa^r$ and real expansion coefficients $p_r$ is defined by the equation
\[  \label{Q1ptriple}
\mathcal{Q}_{1}\left[p \left( \kappa \right)\right]= \frac{\mathcal{Q}_{1}(p_{0})}{3} \left[ e^{\kappa} +2 e^{-\kappa/2}
\cos\left(\frac{\sqrt{3}}{2} \kappa \right) \right],
\]
with
\[\label{SasaQ1poly}
\mathcal{Q}_{1}(p)\equiv \frac{1 }{p-{\rm{i}}\alpha} + \frac{1}{p+{\rm{i}}\alpha}+p,
\]
$p_0=\pm \sqrt{3}/2$, the real vector $\textbf{\emph{s}}=(s_1, s_2, \cdots)$ is defined by the expansion
\begin{equation}
\ln \left[\left(\frac{2p_{0}}{p_{1}\kappa} \right) \left( \frac{ p \left( \kappa \right)-p_{0}}{p \left( \kappa \right)+p_{0}} \right)  \right] = \sum_{r=1}^{\infty}s_{r} \kappa^r,  \label{schurcoeffsr}
\end{equation}
the asterisk `*' represents complex conjugation, and
\[
(a_{1,1}, \cdots, a_{3N_1-1, 1}), \hspace{0.06cm} (a_{1,2}, \cdots, a_{3N_2-2, 2})
\]
are free real constants.
\end{quote}

In these solutions, subscripts of matrix elements in Eq.~(\ref{Blockmatrix}) jump by 3. Thus, the corresponding determinant in Eq.~(\ref{sigmaTheorem3}) is called a determinant of 3-reduced Schur polynomials \cite{KajiOhta1998PIV}.

\textbf{Note 1}\hspace{0.05cm} When we choose $p_0=\sqrt{3}/2$, the first few coefficients of $p_r$, $\beta_{r}$, $\theta_{r}$, and $s_r$ are
\[
p_{1}=\frac{12^{1/6}}{2}, \quad p_2=\frac{12^{-1/6}}{2}, \quad p_3=\frac{1}{4\sqrt{3}},
\]
\[
\beta_1=\frac{9}{8} 12^{1/6},  \quad \beta_2=\frac{9}{8} \cdot \frac{3^{5/6}}{2^{1/3}}, \quad \beta_3=\frac{19 \sqrt{3}}{16},
\]
\[
\theta_1=\frac{12^{1/6}}{\sqrt{3}+{\rm{i}}}, \quad \theta_2=\frac{{\rm{i}}}{12^{1/6} \left(\sqrt{3}+{\rm{i}}\right)^2}, \quad \theta_3=0,
\]
\[
s_1=0, \quad s_2=0, \quad s_3=-\frac{1}{40}.
\]
If we choose $p_0=-\sqrt{3}/2$, then $p_r$ and $\beta_r$ would switch sign, $\theta_r$ change to $\theta_r^*$, and $s_r$ remain the same.

\textbf{Note 2}\hspace{0.05cm} If we choose $p_0=-\sqrt{3}/2$ and keep all internal parameters $(a_{r,1}, a_{r,2})$ unchanged, then the resulting solution $\tilde{u}(x, t)$ would be related to the solution $u(x, t)$ with $p_0=\sqrt{3}/2$ as $\tilde{u}(x, t)=u^*(-x, -t)$.

\textbf{Note 3}\hspace{0.05cm} Internal parameters $a_{3n, 1}$ and $a_{3n,2}$ $(n=1, 2, \cdots)$ do not affect solutions in Theorem~1, for reasons which can be found in \cite{YangYang3wave}. Thus, we will set them as zero in later text.

The simplest solution of this class --- the fundamental rational soliton, is obtained when we set $N_1=0$ and $N_2=1$ in Theorem 1. In this case, the solution has a single real parameter $a_{1,2}$, which can be normalized to zero through a shift of the $x$ axis. The resulting solution, for both $p_0=\pm \sqrt{3}/2$, is
\[ \label{fund}
u_1(x, t)=\hat{u}_1(x, t) e^{{\rm{i}}[\frac{1}{2}(x + 6 t)-\frac{1}{8}t]},
\]
where
\[
\hat{u}_1(x, t)=  \frac{3 \hat{x}^2 + 3 {\rm{i}} \hat{x}-2 }{3\hat{x}^2+1},
\]
and
\[ \label{xhat}
\hat{x}\equiv x+\frac{33}{4}t
\]
is a moving coordinate. The graph of this solution is plotted in Fig.~1. This solution is a rational soliton moving on the constant-amplitude background (\ref{bc}) with velocity $-33/4$. Its 3D graph shows a W-shape along the $\hat{x}$ direction and has sometimes been called a W-shaped rational soliton in the literature \cite{Zhao14,Zhao16}. Its height, i.e., max($|u_1|$), is 2.
\begin{figure}[h]
\begin{center}
\includegraphics[scale=0.255, bb=690 000 285 440]{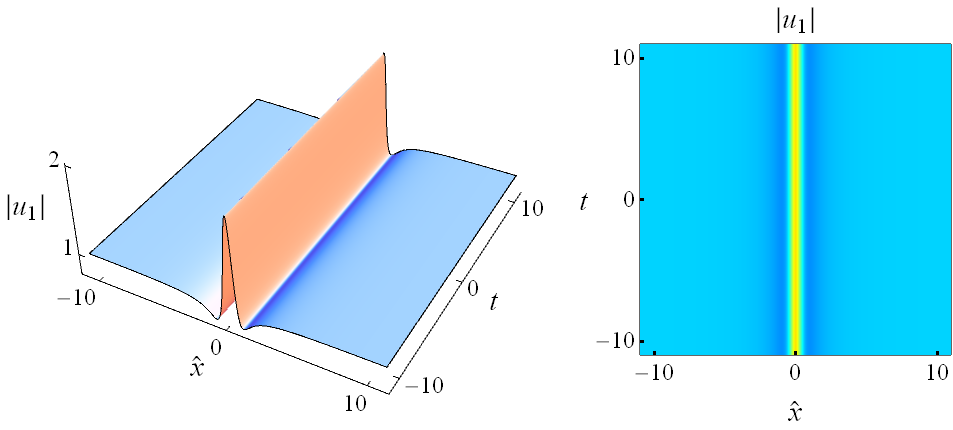}
\caption{Graph of the fundamental rational soliton $|u_1(x, t)|$ in Eq.~(\ref{fund}). Left: 3D plot. Right: density plot. The horizontal axes are $\hat{x}=x+(33/4)t$. }
\end{center}
\end{figure}

\subsection{Generalized Okamoto polynomials}
We will show in later text that rational solutions in Theorem~1 contain partial-rogue waves but are not all partial-rogue waves. The question of what solutions in Theorem~1 are partial-rogue waves turns out to be closely related to root properties of generalized Okamoto polynomials. So, we will introduce these polynomials and examine their root structures next.

Original Okamoto polynomials arose in Okamoto's study of rational solutions to the Painlev\'e IV equation \cite{Okamoto1986}. He showed that a class of such rational solutions can be expressed as the logarithmic derivative of certain special polynomials, which are now called Okamoto polynomials. These original polynomials were later generalized, and the generalized Okamoto polynomials provide a more complete set of rational solutions to the Painlev\'e IV equation \cite{KajiOhta1998PIV,Noumi1999,Clarkson2003PIV,Clarkson2006PIV}. In addition, determinant expressions for the original and generalized Okamoto polynomials were discovered \cite{KajiOhta1998PIV,Noumi1999,Clarkson2003PIV,Clarkson2006PIV}.

Let $p_{j}(z)$ be Schur polynomials defined by
\[ \label{defpkz}
\sum_{j=0}^{\infty}  p_{j}(z) \epsilon^j= \exp \left(z \epsilon + \epsilon^2 \right),
\]
with $p_{j}(z)\equiv 0$ for $j<0$. Then, generalized Okamoto polynomials $Q_{N_1, \hspace{0.04cm} N_2}(z)$, with $N_1, N_2$ being nonnegative integers, are defined as
\[\label{OkamotoPoly1}
Q_{N_1, \hspace{0.04cm} N_2}(z) = \mbox{Wron}[p_2, p_5, \cdots, p_{3N_1-1}, p_1, p_4, \cdots, p_{3N_2-2}],
\]
or equivalently,
\[\label{OkamotoPoly2}
Q_{N_1, \hspace{0.04cm} N_2}(z) =
      \left| \begin{array}{cccc}
         p_{2} & p_{1} & \cdots &  p_{3-N_1-N_2} \\
        \vdots& \vdots & \vdots & \vdots \\
         p_{3N_1-1} & p_{3N_1-2} & \cdots &  p_{2N_1-N_2} \\
        p_{1} & p_{0} & \cdots &  p_{2-N_1-N_2} \\
        \vdots& \vdots & \vdots & \vdots \\
         p_{3N_2-2} & p_{3N_2-3} & \cdots &  p_{2N_2-N_1-1}
       \end{array}
 \right|,
\]
since $p_{j+1}'(z)=p_j(z)$ from the definition of $p_j(z)$ in Eq.~(\ref{defpkz}), where the prime represents differentiation. The first few $Q_{N_1, \hspace{0.04cm} N_2}(z)$ polynomials are
\begin{eqnarray*}
&& Q_{1,0}(z)=\frac{1}{2}(z^2+2), \\
&& Q_{2,0}(z)=\frac{1}{80}(z^6+10 z^4+20 z^2+40), \\
&& Q_{0,1}(z)=z, \\
&& Q_{1,1}(z)=\frac{1}{2}(-z^2+2) \\
&& Q_{2,1}(z)=\frac{1}{20}z(z^4-20),\\
&& Q_{0,2}(z)=\frac{1}{8}(z^4+4 z^2-4), \\
&& Q_{1,2}(z)=\frac{1}{8}(-z^4+4 z^2+4),            \\
&& Q_{2,2}(z)=\frac{1}{80}(-z^6+10 z^4-20 z^2+40).
\end{eqnarray*}
Note that our definition of generalized Okamoto polynomials is different from that by Clarkson in Refs.~\cite{Clarkson2003PIV,Clarkson2006PIV}. Denoting the $Q_{m,\hspace{0.03cm}n}(z)$ polynomial introduced in \cite{Clarkson2003PIV,Clarkson2006PIV} as $Q_{m,\hspace{0.03cm}n}^{[\textrm{C}]}(z)$, then our polynomial $Q_{N_1, \hspace{0.03cm} N_2}(z)$ is related to $Q_{m,\hspace{0.03cm}n}^{[C]}(z)$ as
\begin{equation} \label{conn}
Q_{N_1, N_2}(z)=\left\{\begin{array}{ll}
\gamma_{N_1, \hspace{0.03cm} N_2}^{(1)} Q^{[\textrm{C}]}_{N_2-N_1, \hspace{0.04cm} -N_2}\left(\sqrt{3}\hspace{0.05cm} z/2\right), & N_1\ge  N_2, \\
\gamma_{N_1, \hspace{0.03cm} N_2}^{(2)} Q^{[\textrm{C}]}_{N_2-N_1, \hspace{0.04cm} N_1+1}\left(\sqrt{3}\hspace{0.05cm} z/2\right), & N_1\le N_2,
\end{array}\right.
\end{equation}
where $\gamma_{N_1, \hspace{0.04cm} N_2}^{(1)}$ and $\gamma_{N_1, \hspace{0.04cm} N_2}^{(2)}$ are certain real constants.

Clarkson \cite{Clarkson2003PIV} observed an interesting symmetry relation between $Q_{n,\hspace{0.03cm}m}^{[C]}(z)$ and $Q_{m,\hspace{0.03cm}n}^{[C]}({\rm{i}}z)$ based on examples. Using that symmetry and the above polynomial connection (\ref{conn}), we obtain symmetry relations for our polynomials $Q_{N_1,\hspace{0.03cm} N_2}(z)$ as
\[ \label{Qconnection1}
Q_{N_1, \hspace{0.04cm} N_1-N_2}(z)=b_1 \hspace{0.03cm} e^{-\frac{1}{2} \textrm{i} \pi \hspace{0.03cm} d_{N_1,N_2}} Q_{N_1, \hspace{0.04cm} N_2}({\rm{i}}z), \quad N_1 \ge N_2,
\]
\[ \label{Qconnection2}
Q_{N_2-N_1-1, \hspace{0.04cm} N_2}(z) = b_2 \hspace{0.03cm} e^{-\frac{1}{2} \textrm{i} \pi \hspace{0.03cm} d_{N_1, N_2}} Q_{N_1, \hspace{0.04cm} N_2}({\rm{i}}z),\quad N_1<N_2,
\]
where
\[
d_{N_1, \hspace{0.04cm} N_2}=N_1^2+N_2^2-N_1 N_2 +N_1
\]
is the degree of the $Q_{N_1,N_2}(z)$ polynomial, $b_1=\pm 1$ is the sign of the ratio between coefficients of the highest $z$-power terms in $Q_{N_1,N_2}(z)$ and $Q_{N_1, \hspace{0.04cm} N_1-N_2}(z)$, while $b_2=\pm 1$ is the sign of the ratio between coefficients of the highest $z$-power terms in $Q_{N_1,N_2}(z)$ and $Q_{N_2-N_1-1, \hspace{0.04cm} N_2}(z)$. In the special case of $N_2=0$, the symmetry (\ref{Qconnection1}) further reduces to
\[ \label{Qconnection3}
Q_{N_1,N_1}(z)=Q_{N_1, 0}({\rm{i}}z).
\]

For our partial-rogue wave problem, it turns out from later text that we need generalized Okamoto polynomials which have either real or imaginary roots, but not both. In addition, zero cannot be a root. To identify such polynomials, we plot in Fig.~2 roots of $Q_{N_1, N_2}(z)$ in the complex $z$ plane for $0\le N_1, N_2\le 3$. We can see from this figure that the polynomials that fit our requirements are $Q_{N_1, 0}(z)$ and $Q_{N_1, N_1}(z)$ polynomials, which lie in the first column and on the diagonal of Fig.~2, respectively. The $Q_{N_1, 0}(z)$ polynomials in the first column have only imaginary roots but not real roots. The $Q_{N_1, N_1}(z)$ polynomials on the diagonal have only real roots but not imaginary roots, which is not surprising given the connection between $Q_{N_1, N_1}$ and $Q_{N_1, 0}$ polynomials in Eq.~(\ref{Qconnection3}). For both polynomials, zero is not a root. All other polynomials in Fig.~2 have both real and imaginary roots, and are thus not useful for the partial-rogue problem. We note by passing that these root structures in Fig.~2 are consistent with the two symmetries of generalized Okamoto polynomials in Eqs.~(\ref{Qconnection1})-(\ref{Qconnection2}).

\begin{figure}[h]
\begin{center}
\includegraphics[scale=0.625, bb=140 000 285 400]{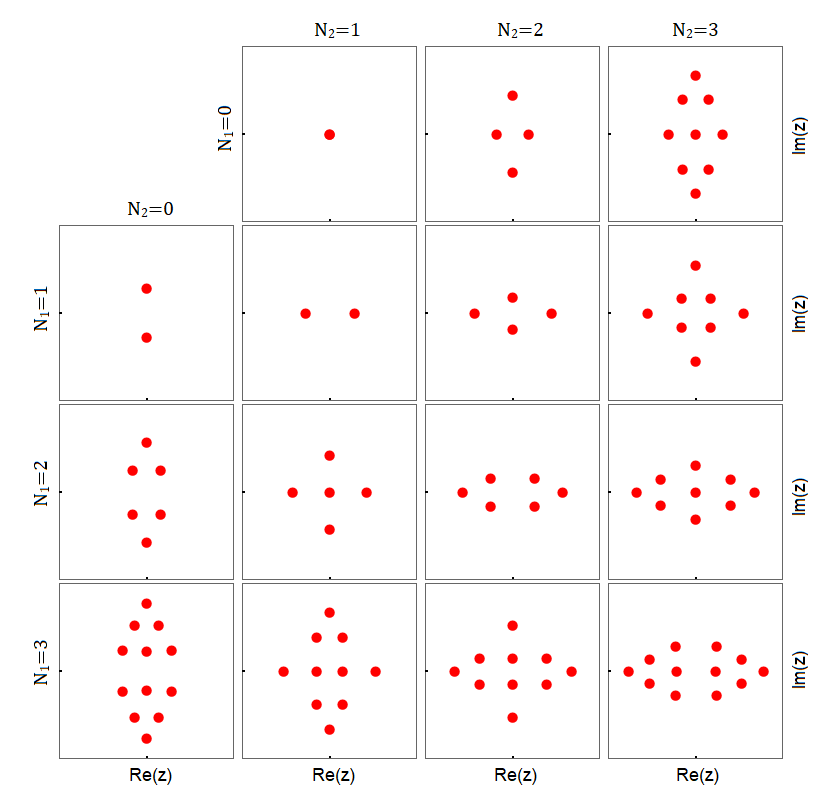}
\caption{Roots of generalized Okamoto polynomials $Q_{N_1, N_2}(z)$ in the complex $z$ plane for $0\le N_1, N_2\le 3$.  In all panels, $-5 \leq \mbox{Re}(z), \mbox{Im}(z) \leq 5$. }
\end{center}
\end{figure}

Multiplicity of these nonzero real or imaginary roots is also important to us. Our numerical checking shows that nonzero roots are all simple for every generalized Okamoto polynomial. This will make our results of the next section a bit simpler.

\section{Partial-rogue waves}
According to our definition, partial-rogue waves are localized waves that ``come from nowhere but leave with a trace". Thus, we impose the following boundary conditions
\[ \label{bc2}
u(x, t) \to e^{{\rm{i}}[\alpha(x + 6 t)-\alpha^3 t]}, \quad t \to  -\infty \hspace{0.08cm} \mbox{or} \hspace{0.1cm} x\to \pm \infty,
\]
where $\alpha=1/2$. In addition, we require $u(x, t)$ not to approach this constant-amplitude background as $t\to +\infty$.

Only a small portion of rational solutions in Theorem~1 are partial-rogue waves. This is not surprising, since the fundamental rational soliton in Fig.~1 is not a partial-rogue wave already. We will show that a rational solution in Theorem~1 is a partial-rogue wave only if the associated generalized Okamoto polynomial has either imaginary or real roots, but not both. This result is summarized in the following two theorems, for imaginary roots and real roots, respectively.

\begin{quote}
\textbf{Theorem 2} \hspace{0.05cm} If the generalized Okamoto polynomial $Q_{N_1, N_2}(z)$ has imaginary but not real roots, and each imaginary root is simple, then the rational solution $u_{N_1, N_2}(x, t)$ in Theorem~1 with $p_0=-\sqrt{3}/2$ is a partial-rogue wave. When $t\gg 1$, this solution approaches the constant-amplitude background $e^{{\rm{i}}[\frac{1}{2}(x + 6 t)-\frac{1}{8} t]}$, except when $x$ is in the $O(1)$ neighborhood of the location
\[ \label{x0hat1}
x_0=-\frac{33}{4}t-{\rm{i}}z_0\frac{3^{3/4}}{2^{1/2}}t^{1/2}+\frac{2}{12^{1/6}}\Delta,
\]
where a fundamental rational soliton $\hat{u}_1(x-x_0, t) e^{{\rm{i}}[\frac{1}{2}(x + 6 t)-\frac{1}{8}t]}$ lies. Here, $z_0$ is each of $Q_{N_1, N_2}(z)$'s imaginary roots, and $\Delta$ is a $z_0$-dependent $O(1)$ quantity whose expression will be given by Eq.~(\ref{Delta}) in later text. The error of this fundamental rational soliton approximation is $O(|t|^{-1/2})$. Expressed mathematically, when $t\gg 1$ and $|x-x_0|=O(1)$,
\[ \label{limit1}
u_{N_1, N_2}(x, t) \to \hat{u}_1(x-x_0, t) e^{{\rm{i}}[\frac{1}{2}(x + 6 t)-\frac{1}{8}t]}+O(|t|^{-1/2}).
\]
When $t\gg 1$ and $|x-x_0| \gg 1$, or when $t\ll -1$,
\[ \label{limit2}
u_{N_1, N_2}(x, t) \to e^{{\rm{i}}[\frac{1}{2}(x + 6 t)-\frac{1}{8}t]}.
\]
\end{quote}

\begin{quote}
\textbf{Theorem 3} \hspace{0.05cm} If the generalized Okamoto polynomial $Q_{N_1, N_2}(z)$ has real but not imaginary roots, and each real root is nonzero and simple, then the rational solution $u_{N_1, N_2}(x, t)$ in Theorem~1 with $p_0=\sqrt{3}/2$ is a partial-rogue wave. When $t\gg 1$, this solution approaches the constant-amplitude background $e^{{\rm{i}}[\frac{1}{2}(x + 6 t)-\frac{1}{8} t]}$, except when $x$ is in the $O(1)$ neighborhood of the location
\[ \label{defx02}
x_0=-\frac{33}{4}t+z_0\frac{3^{3/4}}{2^{1/2}}t^{1/2}-\frac{2}{12^{1/6}}\Delta,
\]
where a fundamental rational soliton $\hat{u}_1(x-x_0, t) e^{{\rm{i}}[\frac{1}{2}(x + 6 t)-\frac{1}{8}t]}$ lies. Here, $z_0$ is each of $Q_{N_1, N_2}(z)$'s real roots, and $\Delta$ is a $z_0$-dependent $O(1)$ quantity whose expression will be given by Eq.~(\ref{Delta}) in later text. The error of this fundamental rational soliton approximation is $O(|t|^{-1/2})$. Mathematical expressions of these results are the same as in Eqs.~(\ref{limit1})-(\ref{limit2}) of Theorem 2, except for the different formula (\ref{defx02}) for the soliton location $x_0$.
\end{quote}

Proofs of these two theorems will be given in the next section.

These two theorems, together with root structures of generalized Okamoto polynomials in Fig.~2, predict that rational solutions $u_{N_1, 0}(x,t)$ $(N_1\ge 1)$ with $p_0=-\sqrt{3}/2$, as well as $u_{N_1, N_1}(x, t)$ $(N_1\ge 1)$ with $p_0=\sqrt{3}/2$, are partial-rogue waves. As $t\to -\infty$, they approach the constant-amplitude background. As $t\to +\infty$, they split into several fundamental rational solitons, and the number of such fundamental solitons is equal to the number of real or imaginary roots in the underlying generalized Okamoto polynomial. These results are not dependent on values of internal parameters $a_{r,1}, a_{r,2}$ $(r=1, 2, \cdots)$. This is not surprising, since when $|t|\gg 1$, those internal parameters in the solution will play a less significant role.

Next, we numerically verify these two theorems. In these numerical verifications, we will set $a_{1,1}=0$ through a shift of the $x$ axis.

First, we consider Theorem~2. Based on Fig.~2, this theorem predicts that rational solutions $u_{1, 0}$, $u_{2, 0}$, and $u_{3, 0}$, with $p_0=-\sqrt{3}/2$, are partial-rogue waves. To verify this, we take internal parameters in these three solutions respectively as
\begin{eqnarray}
&& \hspace{-0.7cm} a_{2,1}=0, \label{para1} \\
&& \hspace{-0.7cm} a_{2,1}=0, \hspace{0.07cm} a_{4,1}=2, \hspace{0.07cm} a_{5,1}=-3, \label{para2} \\
&& \hspace{-0.7cm} a_{2,1}=0, \hspace{0.07cm} a_{4,1}=2, \hspace{0.07cm} a_{5,1}=-3, \hspace{0.07cm} a_{7,1}=a_{8,1}=0. \label{para3}
\end{eqnarray}
The corresponding true solutions are plotted from Theorem~1 and displayed in Fig.~3. As can be seen, these are indeed partial-rogue waves that arise from the constant background but do not retreat back to it, in agreement with Theorem~2.

\begin{figure}[h]
\begin{center}
\includegraphics[scale=0.46, bb=275 000 285 210]{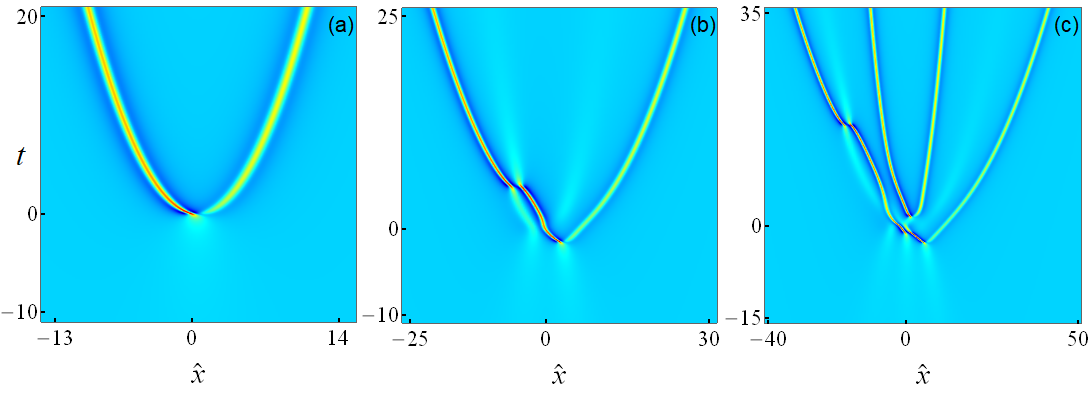}
\caption{Density plots of partial-rogue waves $|u_{1, 0}(x,t)|$ (a), $|u_{2, 0}(x, t)|$ (b), and $|u_{3, 0}(x, t)|$ (c), with $p_0=-\sqrt{3}/2$ and internal parameter values in Eqs.~(\ref{para1}), (\ref{para2}), and (\ref{para3}), respectively. The horizontal axes are $\hat{x}=x+(33/4)t$. }
\end{center}
\end{figure}

Theorem~2 also predicts that, as $t\to+\infty$, these partial-rogue waves would split into several  fundamental rational solitons. Fig.~3 confirms that this is indeed the case. The reader may notice that individual fundamental solitons at large time in Fig.~3 appear to have different heights, while Theorem~2 predicts these fundamental solitons should approach the same height. It turns out that this discrepancy is due to the fact that the time shown in Fig.~3 is not large enough. We have checked that as time increases further, all these humps indeed approach the same height 2, which is the height of the fundamental rational soliton (\ref{fund}). To demonstrate, we choose the $|u_{2, 0}(x, t)|$ solution in Fig.~3(b), and track the heights of its two humps versus time. The corresponding graphs are plotted in the left panel of Fig.~4. The height 2 of the fundamental soliton is also shown for comparison. One can see that the heights of both humps monotonically approach the height of the fundamental soliton as $t\to+\infty$, in agreement with Theorem~2.

\begin{figure}[h]
\begin{center}
\includegraphics[scale=0.38, bb=360 000 285 330]{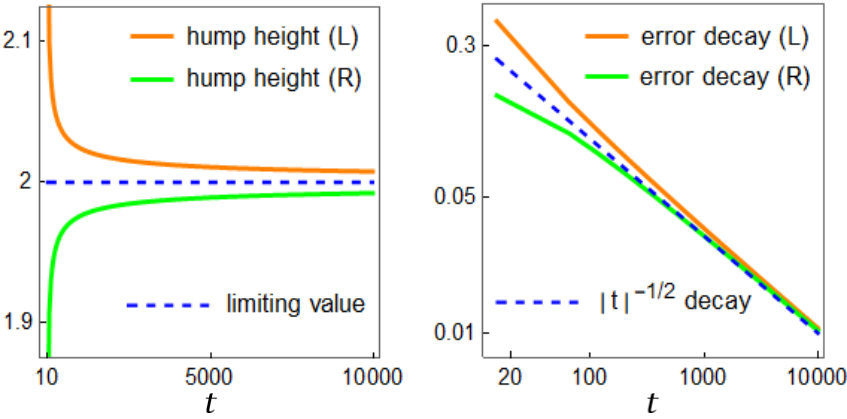}
\caption{Quantitative comparison between the true partial-rogue solution of Fig.~3(b) and its prediction from Theorem~2. Left: graphs of the two humps' heights versus time (upper one for the left hump and lower one for the right hump); the theoretical limiting value of 2 is also shown (as dashed line) for comparison. Right: errors versus time for predicted locations of the two humps at large time; the $|t|^{-1/2}$ decay is also plotted for comparison.}
\end{center}
\end{figure}

To show further quantitative comparison, we again choose the $|u_{2, 0}(x, t)|$ solution in Fig.~3(b). This time, we track true locations of its two humps at each large time, and compare them to predicted locations (\ref{x0hat1}) in Theorem~2. The errors of these predictions, defined as the absolute difference between true and predicted hump locations, versus time are plotted in the right panel of Fig.~4. This panel shows that the errors decay at the rate of $O(|t|^{-1/2})$, which matches our error estimate in the asymptotics (\ref{limit1}). Thus, Theorem~2 is fully confirmed.

Next, we numerically confirm Theorem~3. Based on Fig.~2, this theorem predicts that rational solutions $u_{1,1}(x,t)$, $u_{2,2}(x,t)$, and $u_{3,3}(x,t)$ with with $p_0=\sqrt{3}/2$ are partial-rogue waves. To verify this, we take internal parameters in these three solutions respectively as
\begin{eqnarray}
&& \hspace{-0.8cm} a_{2,1}=0, \hspace{0.07cm} a_{1,2}=3,  \label{para1b} \\
&& \hspace{-0.8cm} a_{2,1}=a_{4,1}=a_{5,1}=0, \hspace{0.07cm} a_{1,2}=a_{2,2}=a_{4,2}=3, \label{para2b} \\
&& \hspace{-0.8cm} a_{2,1}=a_{4,1}=a_{5,1}=a_{7,1}=a_{8,1}=0, \nonumber \\
&& \hspace{-0.8cm} a_{1,2}=a_{2,2}=a_{4,2}=a_{5,2}=a_{7,2}=3. \label{para3b}
\end{eqnarray}
The corresponding true solutions are plotted from Theorem~1 and displayed in Fig.~5. We can see that these are indeed partial-rogue waves, in agreement with Theorem~3. We have also done quantitative comparison between these true partial-rogue waves and their theoretical predictions in Theorem~3, similar to what we did in Fig.~4. That comparison also confirmed Theorem~3 quantitatively. Details will be omitted for brevity.

\begin{figure}[h]
\begin{center}
\includegraphics[scale=0.46, bb=260 000 285 210]{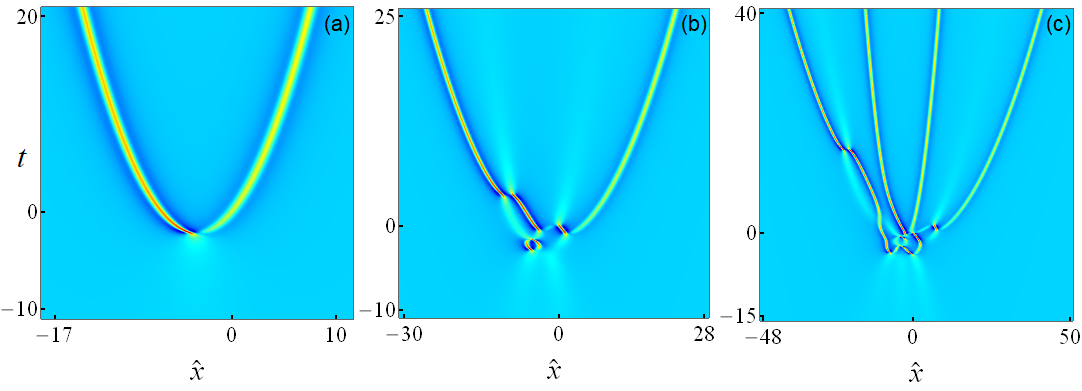}
\caption{Density plots of partial-rogue waves $|u_{1, 1}(x,t)|$ (a), $|u_{2, 2}(x, t)|$ (b), and $|u_{3, 3}(x, t)|$ (c), with $p_0=\sqrt{3}/2$ and internal parameter values in Eqs.~(\ref{para1b}), (\ref{para2b}), and (\ref{para3b}), respectively. The horizontal axes are $\hat{x}=x+(33/4)t$. }
\end{center}
\end{figure}

\section{Proofs of theorems}
The proofs of Theorems 2 and 3 follow the asymptotic analysis we developed in \cite{YangYang2021a,YangYang2021b,YangYang2022} for rogue patterns in integrable systems and lump patterns in the Kadomtsev-Petviashvili I equation. We will only prove Theorem~2, since the proof of Theorem~3 is very similar.

We first rewrite $\sigma_{k,l}$ in Eq.~(\ref{sigmaTheorem3}) as a larger determinant with simpler matrix elements \cite{OhtaJY2012,YangYang2021a}
\[ \label{3Nby3Ndet}
\sigma_{k,l}=\left|\begin{array}{cc}
\textbf{O}_{N\times N} & \Phi_{N\times \widehat{N}} \\
-\Psi_{\widehat{N}\times N} & \textbf{I}_{\widehat{N}\times \widehat{N}} \end{array}\right|,
\]
where $N=N_1+N_2$, $\widehat{N}=\mbox{max}(3N_1, 3N_2-1)$,
\begin{eqnarray*}
&&\hspace{-0.5cm} \Phi_{i,j}^{(k,l)}=\left\{\begin{array}{ll}
h_0^{j-1} S_{3i-j} \left(\textbf{\emph{x}}^{+}_1(k,l) + (j-1) \textbf{\emph{s}} \right), & i\le N_1,  \\
h_0^{j-1} S_{3(i-N_1)-j-1} \left(\textbf{\emph{x}}^{+}_2(k,l) + (j-1) \textbf{\emph{s}} \right), & i> N_1, \end{array} \right.   \\
&&\hspace{-0.5cm} \Psi_{i,j}^{(k,l)}=\left\{\begin{array}{ll}
h_0^{i-1} S_{3j-i} \left(\textbf{\emph{x}}^{-}_1(k,l) + (i-1) \textbf{\emph{s}} \right), & j\le N_1,  \\
h_0^{i-1} S_{3(j-N_1)-i-1} \left(\textbf{\emph{x}}^{-}_2(k,l) + (i-1) \textbf{\emph{s}} \right), & j>N_1, \end{array} \right.
\end{eqnarray*}
and $h_0\equiv p_1/2p_0$. Next, we apply the Laplace expansion to Eq.~(\ref{3Nby3Ndet}) and get
\begin{eqnarray} \label{sigmaLap}
&& \hspace{-0.8cm} \sigma_{k,l}=\sum_{0\leq\nu_{1} < \nu_{2} < \cdots < \nu_{N}\leq \widehat{N}-1}
\det_{1 \leq i, j\leq N} \Phi_{i, \nu_j}^{(k,l)} \times \det_{1 \leq i, j\leq N}\Psi_{i, \nu_j}^{(k,l)}. \hspace{0.65cm}
\end{eqnarray}
To analyze $\sigma_{k,l}$'s large-time behavior, we need large-time asymptotics of $S_j(\textbf{\emph{x}}^{\pm}_I(k,l) + \nu \textbf{\emph{s}})$. Notice that for $|x|\gg 1$ and $|t|\gg 1$, \[
x_{1,I}^+(k,l) \sim p_1(x+6t)+\beta_1 t =p_1\hat{x},
\]
where $\hat{x}$ is as defined in (\ref{xhat}), and $I=1, 2$. Similarly,
\begin{eqnarray}
x_{2,I}^+(k,l) \sim p_2 \hat{x}+\mu_2 t, \quad x_{3,I}^+(k,l) \sim p_3 \hat{x}+\mu_3 t,
\end{eqnarray}
where
\[
\mu_2=9p_2/2, \quad \mu_3=12 p_3.
\]
Thus, when $|t|\gg 1$ and $\hat{x}=O(|t|^{1/2})$, we have the following leading-order asymptotics
\[ \label{Skasym1}
S_{j}\left(\textbf{\emph{x}}_I^{+}(k,l) + \nu  \textbf{\emph{s}} \right) \sim S_j(\textbf{v}),
\]
where
\[  \label{vdef}
\textbf{v}=\left(p_1\hat{x}, \hspace{0.04cm} \mu_2 t,  \hspace{0.04cm} 0, \hspace{0.04cm} 0,  \cdots\right).
\]
By comparing the definition of Schur polynomials $S_j(\textbf{v})$ in (\ref{Elemgenefunc}) to the definition of $p_j(z)$ polynomials in (\ref{defpkz}), we see that
\begin{equation} \label{Skorder}
S_j(\textbf{v})=\left(\mu_2t\right)^{j/2}\hspace{-0.1cm} p_{j}(z),
\end{equation}
where
\[ \label{defz1}
z\equiv \frac{\hspace{0.06cm} p_1\hat{x}}{\sqrt{\mu_2 t}}.
\]
Using these results and similar ones for $S_{j}(\textbf{\emph{x}}_I^{-}(k,l) + \nu  \textbf{\emph{s}})$, we find that the leading-order term of $\sigma_{k,l}$ in Eq.~(\ref{sigmaLap}) is
\[ \label{sigmaasym}
\sigma_{k,l} \sim h_0^{2m_0} (\mu_2 t)^{n_0} Q^2_{N_1, N_2}(z), \quad |t|\gg 1,
\]
where $m_0$ and $n_0$ are certain positive integers. Since $p_0<0$ in Theorem 2, $\mu_2<0$. Thus, for large negative time, $z$ in Eq.~(\ref{defz1}) is real. Then, Eq.~(\ref{sigmaasym}) tells us that, if $Q_{N_1, N_2}(z)$ does not have real roots as assumed in Theorem~2, the above leading-order asymptotics for $\sigma_{k,l}$ would not vanish. Since this asymptotics is independent of $(k,l)$, then $\sigma_{1,0}/\sigma_{0,0}$ would approach 1 when $t\to -\infty$, which means that
\[ \label{uasym5}
u_{N_1, N_2}(x, t)\to e^{{\rm{i}}[\frac{1}{2}(x + 6 t)-\frac{1}{8} t]}, \quad t\to -\infty,
\]
in view of Eq.~(\ref{Schpolysolu1}).

When $t\gg 1$, $z$ in Eq.~(\ref{defz1}) is imaginary. If this $z$ value is not near an imaginary root $z_0$ of the $Q_{N_1, N_2}(z)$ polynomial, i.e., $|\hat{x}-\hat{x}_0|\gg 1$, where
\[ \label{x0hat}
\hat{x}_0=\frac{\sqrt{\mu_2t}}{\hspace{0.06cm} p_1}z_0=-{\rm{i}}z_0\frac{3^{3/4}}{2^{1/2}}t^{1/2},
\]
then the leading-order asymptotics (\ref{sigmaasym}) does not vanish either. For similar reasons as above, $u_{N_1, N_2}(x, t)$ would approach the background $e^{{\rm{i}}[\frac{1}{2}(x + 6 t)-\frac{1}{8} t]}$ as well.

When $t\gg 1$ and $|\hat{x}-\hat{x}_0|=O(1)$, the $z$ value from Eq.~(\ref{defz1}) is near $z_0$, and the leading-order asymptotics (\ref{sigmaasym}) breaks down. In this case, a more refined asymptotic analysis is needed. The starting point is a more refined asymptotics for $S_{j}\left(\textbf{\emph{x}}_I^{+}(k,l) + \nu  \textbf{\emph{s}} \right)$,
\[
S_{j}\left(\textbf{\emph{x}}_I^{+}(k,l) + \nu  \textbf{\emph{s}} \right)=
S_{j}\left(\hat{\textbf{v}}_I\right) \left(1+O(|t|^{-1})\right),
\]
where
\[
\hat{\textbf{v}}_I=\left(x_{1,I}^{+}(k,l), p_2\hat{x}+\mu_2t, \mu_3t, 0, 0, \cdots\right).
\]
Here, the fact of $s_1=0$ has been utilized. Let us split $\hat{\textbf{v}}_1$ and $\hat{\textbf{v}}_2$ as
\begin{eqnarray}
&& \hat{\textbf{v}}_1=\textbf{w}+\left(0, p_2\hat{x}, \mu_3 t, 0, 0, \cdots\right), \\
&& \hat{\textbf{v}}_2=\textbf{w}+\left(a_{1,2}-a_{1,1}, p_2\hat{x}, \mu_3 t, 0, 0, \cdots\right),
\end{eqnarray}
where
\[
\textbf{w}\equiv \left(x_{1,1}^{+}(k,l), \mu_2t, 0, 0, \cdots \right).
\]
Then, using the definition (\ref{Elemgenefunc}) of Schur polynomials, we can readily find that
\begin{eqnarray}
&& \hspace{-1.1cm} S_{j}\left(\textbf{\emph{x}}_1^{+}(k,l) + \nu  \textbf{\emph{s}} \right)=\left[
S_{j}(\textbf{w})+p_2\hat{x}_0 \hspace{0.03cm} S_{j-2}(\textbf{w})+\mu_3t \hspace{0.04cm} S_{j-3}(\textbf{w})\right] \nonumber \\
&& \hspace{1.5cm} \times \left(1+O(|t|^{-1})\right),    \label{Sjxp1} \\
&& \hspace{-1.1cm} S_{j}\left(\textbf{\emph{x}}_2^{+}(k,l) + \nu  \textbf{\emph{s}} \right)=\left[
S_{j}(\textbf{w})+(a_{1,2}-a_{1,1})S_{j-1}(\textbf{w}) \right.  \nonumber \\
&& \hspace{-0.3cm} \left. +p_2\hat{x}_0 \hspace{0.03cm} S_{j-2}(\textbf{w})+\mu_3t \hspace{0.04cm} S_{j-3}(\textbf{w})\right]\times \left(1+O(|t|^{-1})\right), \label{Sjxp2}
\end{eqnarray}
where
\begin{eqnarray} \label{Skorder2}
&& \hspace{-0.7cm} S_j(\textbf{w})=\left(\mu_2t\right)^{j/2}\hspace{-0.1cm} p_{j}\left(\frac{x_{1,1}^{+}(k,l)}{\sqrt{\mu_2t}}\right)   \nonumber \\
&& \hspace{-0.7cm} =\left(\mu_2t\right)^{j/2}\hspace{-0.1cm} p_{j}\left(z_0+\frac{p_1(\hat{x}-\hat{x}_0)+
k\theta_{1}+ l \theta_{1}^*+ a_{1,1}}{\sqrt{\mu_2t}}\right). \hspace{0.25cm}
\end{eqnarray}
Similar asymptotics can be obtained for $S_{j}\left(\textbf{\emph{x}}_I^{-}(k,l) + \nu  \textbf{\emph{s}} \right)$.

Now, we use these refined asymptotics of $S_{j}\left(\textbf{\emph{x}}_I^{\pm}(k,l) + \nu  \textbf{\emph{s}} \right)$ to determine the leading-order asymptotics of $\sigma_{k,l}$ from Eq.~(\ref{sigmaLap}). This leading-order asymptotics comes from two index-vector contributions, one being $\nu=(0, 1, 2, \cdots, N-2, N-1)$, and the other being $\nu=(0, 1, 2, \cdots, N-2, N)$. For the first index vector, there are two sources of contributions to $\det_{1 \leq i, j\leq N} \Phi_{i, \nu_j}^{(k,l)}$. One is when the $S_{j}(\textbf{w})$ term in (\ref{Sjxp1})-(\ref{Sjxp2}) is chosen in each $\Phi_{i, \nu_j}^{(k,l)}$ element. In view of Eq.~(\ref{Skorder2}), this part of the contribution amounts to
\[
h_0^{m_0}(\mu_2 t)^{(n_0-1)/2}\left(p_1(\hat{x}-\hat{x}_0)+k\theta_{1}+ l \theta_{1}^*+ a_{1,1}\right)Q'_{N_1, N_2}(z_0),   \nonumber
\]
where $m_0$ and $n_0$ are the same as those in Eq.~(\ref{sigmaasym}). The other source of contributions to $\det_{1 \leq i, j\leq N} \Phi_{i, \nu_j}^{(k,l)}$ comes from taking the $S_{j}(\textbf{w})$ term of (\ref{Sjxp1})-(\ref{Sjxp2}) in all columns of the $\Phi_{i, \nu_j}^{(k,l)}$ matrix, except for a single column where the $S_{j-1}(\textbf{w})$, $S_{j-2}(\textbf{w})$, and $S_{j-3}(\textbf{w})$ terms of (\ref{Sjxp1})-(\ref{Sjxp2}) are chosen. Recalling the $\hat{x}_0$ formula (\ref{x0hat}), this part of the contribution amounts to
\begin{eqnarray*}
&& \hspace{-0.5cm} h_0^{m_0}(\mu_2 t)^{(n_0-1)/2}\sum_{j=1}^{N}\left[
(a_{1,2}-a_{1,1})Q_j^{(1)}(z_0)+\frac{p_2}{p_1}z_0Q_j^{(2)}(z_0)  \right. \\
&& \hspace{2.0cm} \left.+\frac{\mu_3}{\mu_2}Q_j^{(3)}(z_0)\right],
\end{eqnarray*}
where $Q_j^{(1)}(z)$ is the $Q_{N_1,N_2}(z)$ determinant (\ref{OkamotoPoly2}), but with its $j$-th column modified so that its first $N_1$ elements become zero, and its remaining elements are the original $p_n(z)$'s with their indices $n$ reduced by one each, and $Q_j^{(2)}(z)$, $Q_j^{(3)}(z)$ are the $Q_{N_1,N_2}(z)$ determinants (\ref{OkamotoPoly2}) but with $p_n(z)$ indices $n$ of their $j$-th column reduced by two and three, respectively. Contributions to $\det_{1 \leq i, j\leq N} \Psi_{i, \nu_j}^{(k,l)}$ of (\ref{sigmaLap}) can be obtained similarly.

For the second index vector of $\nu=(0, 1, 2, \cdots, N-2, N)$ in Eq.~(\ref{sigmaLap}), leading-order contributions to $\det_{1 \leq i, j\leq N} \Phi_{i, \nu_j}^{(k,l)}$ only come from choosing
the $S_{j}(\textbf{w})$ term of (\ref{Sjxp1})-(\ref{Sjxp2}) in each $\Phi_{i, \nu_j}^{(k,l)}$ element, and this contribution amounts to
\[
h_0^{m_0-1}(\mu_2 t)^{(n_0-1)/2}Q'_{N_1, N_2}(z_0)     \nonumber
\]
in view of the relation $p_{j+1}'(z)=p_j(z)$. A similar result can be obtained for $\det_{1 \leq i, j\leq N} \Psi_{i, \nu_j}^{(k,l)}$.

Collecting these results, we find that the leading-order contribution to $\sigma_{k,l}$ in Eq.~(\ref{sigmaLap}) is
\begin{eqnarray}  \label{sigmalead}
&& \hspace{-1.2cm} \sigma_{k,l}\sim
h_0^{2m_0}(\mu_2 t)^{n_0-1}Q'^2_{N_1, N_2}(z_0) \left[\left(p_1(\hat{x}-\hat{x}_0)+k\theta_{1}+ l \theta_{1}^*+ \Delta \right)  \right. \nonumber \\
&& \hspace{-0.3cm} \left. \times \left(p_1(\hat{x}-\hat{x}_0)- k\theta_{1}^*- l \theta_{1}+\Delta \right)+h_0^{-2}\right],
\end{eqnarray}
where
\[ \label{Delta}
\Delta=a_{1,1}+\frac{\sum_{j=1}^{N}\left[
\hat{a}_1 Q_j^{(1)}(z_0)+\frac{p_2}{p_1}z_0Q_j^{(2)}(z_0)+\frac{\mu_3}{\mu_2}Q_j^{(3)}(z_0)\right]}{Q'_{N_1,N_2}(z_0)},
\]
and $\hat{a}_1\equiv a_{1,2}-a_{1,1}$. Since the imaginary root $z_0$ of $Q_{N_1, N_2}(z)$ is simple according to our assumption, $Q'_{N_1, N_2}(z_0)\ne 0$. Thus, the above leading-order asymptotics of $\sigma_{k,l}$ does not vanish. This asymptotics, when inserted into Eq.~(\ref{Schpolysolu1}), gives a fundamental rational soliton $\hat{u}_1(x-x_0, t) e^{{\rm{i}}[\frac{1}{2}(x + 6 t)-\frac{1}{8}t]}$, where the soliton position $x_0$ can be obtained from $\hat{x}_0$ and $\Delta$ as
\[
x_0=-\frac{33}{4}t+\hat{x}_0-\frac{\Delta}{p_1},
\]
which is the same as Eq.~(\ref{x0hat1}) in Theorem~2. The relative error of the leading-order asymptotics (\ref{sigmalead}) of $\sigma_{k,l}$ is $O(|t|^{-1/2})$, which leads to an error of $O(|t|^{-1/2})$ in the above fundamental-soliton approximation. This completes the proof of Theorem~2.

\section{Other types of rational solutions}
Rational solutions in Theorem~1 also contain other types of solutions. One other type is
``waves that come from somewhere but leave without a trace" --- the opposite of partial-rogue waves we considered earlier in this paper. Such solutions obviously exist, because the Sasa-Satsuma equation is invariant under the $(x, t)\to (-x, -t)$ transformation. Thus, from every partial-rogue wave, we can get such a new wave. This way of getting such new solutions will change the background condition (\ref{bc}) though. If we want to preserve that background, then we can just switch the sign of $p_0$ in Theorems~2 and 3, and the resulting solution would be ``waves that come from somewhere but leave without a trace" instead of partial-rogue waves. As an example, we show in Fig.~6(a) such a solution by switching the sign of $p_0$ in the partial-rogue wave of Fig.~3(b). It is noted that a simpler solution of this type has been reported earlier in \cite{LingSS2016} (see Fig.~7(b) there). This type of solutions, although different, are closely related to partial-rogue waves. Thus, it is reasonable for us to call them partial-rogue waves as well for simplicity.

These partial-rogue waves are obtained when the associated generalized Okamoto polynomials have real but not imaginary roots, or imaginary but not real roots. As one can see from Fig.~2, most generalized Okamoto polynomials are not like that. For such polynomials, the associated rational solutions in Theorem~1 would not be partial-rogue waves. Instead, they would be solutions which split into several fundamental (or lower-order) rational solitons as time approaches both $\pm \infty$. As an example, we choose $(N_1, N_2)=(3, 1)$ and $p_0=\sqrt{3}/2$. The corresponding $Q_{3,1}(z)$ polynomial has four nonzero real roots and two imaginary roots, all of which are simple, see Fig.~2. Thus, a simple extension of our earlier asymptotic analysis predicts that, as $t$ approaches $-\infty$, this solution would split into two fundamental rational solitons, but as $t$ approaches $+\infty$, it would split into four fundamental rational solitons. To illustrate, we choose all internal parameters $a_{r,1}, a_{r,2}$ as zero. The resulting true solution from Theorem~1 is displayed in Fig.~6(b). This solution shows that, out of the interaction and collision of two fundamental rational solitons, four fundamental rational solitons emerge. This phenomenon is unusual and fascinating.

\begin{figure}[h]
\begin{center}
\includegraphics[scale=0.26, bb=485 000 285 410]{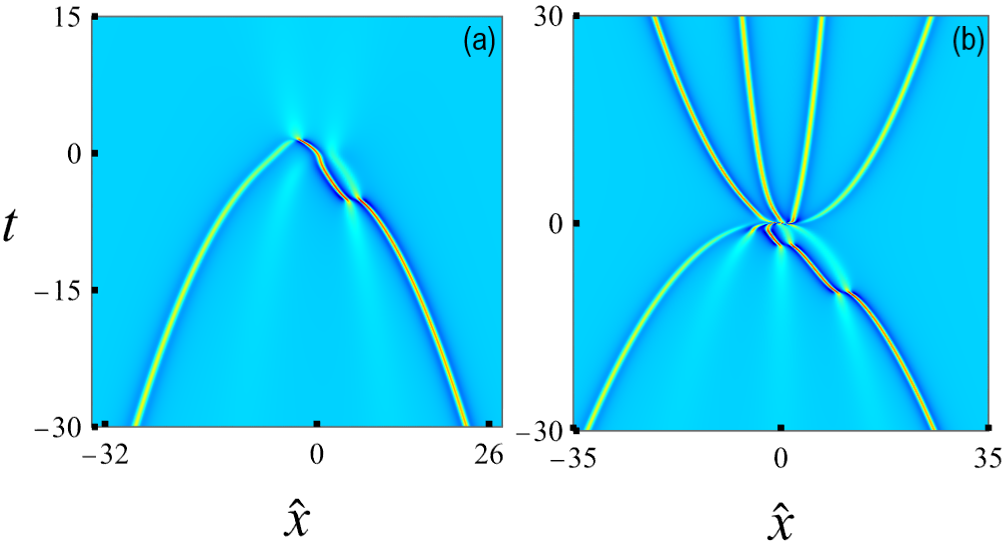}
\caption{Other types of rational solutions in Theorem~1. (a) A wave that comes from somewhere but leaves without a trace. This solution is obtained from $|u_{2, 0}(x, t)|$ of Fig.~3(b), but with its $p_0$ value flipped from $-\sqrt{3}/2$ to $\sqrt{3}/2$.
(b) A wave that comes and leaves with traces. This is the $|u_{3,1}(x,t)|$ solution with $p_0=\sqrt{3}/2$ and all internal parameters as zero. In both panels, the horizontal axes are $\hat{x}=x+(33/4)t$.}
\end{center}
\end{figure}

\section{Summary}
In this article, we have asymptotically and numerically studied partial-rogue waves in the Sasa-Satsuma equation. We have shown that, among a class of rational solutions in this equation that can be expressed through determinants of 3-reduced Schur polynomials, partial-rogue waves would appear if these rational solutions are of certain orders, whose associated generalized Okamoto polynomials have real but not imaginary roots, or imaginary but not real roots. We have further shown that, these partial-rogue waves asymptotically approach the constant-amplitude background as time goes to negative infinity, but split into several fundamental rational solitons as time goes to positive infinity. Our asymptotic predictions are compared to true solutions both qualitatively and quantitatively, and excellent agreement has been obtained.

In earlier work \cite{YangYang2021a,YangYang2021b,YangYang2022}, we linked rogue and lump patterns in the space-time plane to root structures of certain special polynomials in the complex plane. In that work, all roots of the special polynomials contributed to the space-time patterns of solutions. A distinctive feature of our present work is that, the question of partial-rogue waves and their large-time behaviors is linked to only real and imaginary roots of the underlying special polynomials (i.e., generalized Okamoto polynomials). Other complex roots of these polynomials are irrelevant. This feature vaguely resembles an earlier work in \cite{MillerGordon}, where superluminal kinks in the semiclassical sine-Gordon equation were linked to real roots of Yablonskii-Vorob'ev polynomials. This wide variety of connections between nonlinear wave dynamics and certain types of roots in special polynomials is a remarkable phenomenon, and it reflects the richness of wave behaviors in nonlinear partial differential equations.

From a broader perspective, some other solutions are also related to partial-rogue waves. For example, in a two-dimensional multi-component long-wave-short-wave interaction system \cite{He2022}, some solutions describing a resonant collision between lumps and homoclinic orbits are such that the underlying lumps do not exist at large negative time but arise and persist at large positive time. But such solutions may not be called partial-rogue waves since they are not localized in space at intermediate times due to the nonlocal homoclinic-orbit component.

\section*{Acknowledgment}
The work of B.Y. was supported in part by the National Natural Science Foundation of China
(Grant No.12201326), and the work of J.Y. was supported in part by the National Science Foundation (U.S.) under award number DMS-1910282.

\begin{center}
\textbf{Appendix}
\end{center}
\renewcommand{\theequation}{A.\arabic{equation}}

In this appendix, we briefly derive rational solutions given in Theorem~1.

It has been shown in \cite{Feng22b} that the Sasa-Satsuma equation (\ref{SS}) under boundary conditions (\ref{bc}) admits the following solutions
\[
u(x, t)=e^{{\rm{i}} [\alpha(x + 6t)-\alpha^3 t]}\left.\frac{\tau_{1,0}}{\tau_{0,0}}
\right|_{y=r=s=0},
\]
where
\[ \label{taublock}
\tau_{k,l} = \det \left( \begin{array}{cc}
                           \tau_{k,l}^{\left[1,1\right]} & \tau_{k,l}^{\left[1,2\right]} \\
                           \tau_{k,l}^{\left[2,1\right]} & \tau_{k,l}^{\left[2,2\right]}
                         \end{array}
\right),
\]
\[
\tau^{[I, J]}_{k,l}=
\left(
\phi_{i_\nu^{[I]},j_\mu^{[J]}}^{(k,l \hspace{0.04cm} I, J)}
\right)_{1\leq \nu \leq N_{I}, \, 1\leq \mu \leq N_{J}},
\]
$N_1$ and $N_2$ are arbitrary non-negative integers, $\left(i_1^{[I]},i_2^{[I]},\cdots,i_{N_I}^{[I]}\right)$ and $\left(j_1^{[J]},j_2^{[J]},\cdots,j_{N_J}^{[J]}\right)$ are arbitrary sequences of non-negative indices,
\[\label{Sasamijn}
\phi_{i,j}^{(k,l,I,J)}=\left. \mathcal{A}_i \mathcal{B}_j \phi^{(k,l,I,J)}\right|_{p=q, \hspace{0.05cm} \xi_{0,I}(p)=\eta_{0,I}(q), \hspace{0.05cm} \xi_{0,J}(p)=\eta_{0,J}(q)},
\]
\[
\phi^{(k,l,I,J)}=\frac{1}{p+q} \left(-\frac{p-{\rm{i}}\alpha}{q+{\rm{i}}\alpha}\right)^k \left(-\frac{p+{\rm{i}}\alpha}{q-{\rm{i}}\alpha}\right)^l e^{\xi_I(p)+\eta_J(q)},
\]
\[
\xi_I(p)=px+p^2 y + p^3 t+ \frac{1}{p-{\rm{i}}\alpha}r+\frac{1}{p+{\rm{i}}\alpha}s+\xi_{0,I}(p),
\]
\[
\eta_J(q)=qx -q^2 y + q^3 t+ \frac{1}{q+{\rm{i}}\alpha}r+\frac{1}{q-{\rm{i}}\alpha}s+\eta_{0,J}(q),
\]
$p$ is an arbitrary real number, $\xi_{0,1}(p)$ and $\xi_{0,2}(p)$ are arbitrary real functions of $p$, $\mathcal{A}_i$ and $\mathcal{B}_j$ are differential operators
\[
\mathcal{A}_i =\frac{1}{i!}\left[f_1(p)\partial_p\right]^{i},\ \ \ \mathcal{B}_j=\frac{1}{j!}\left[f_2(q)\partial_q\right]^{j},
\]
and $f_{1}(p)$, $f_{2}(q)$ are arbitrary real functions, if the above $\tau_{k,l}$ satisfies the dimension reduction condition
\[ \label{dimred}
\left(\partial_{r} + \partial_{s} + \partial_{x}\right)\tau_{k,l}=C \tau_{k,l},
\]
where $C$ is some constant. The above result can be made even more general by allowing each of $p$ and $q$ to take different values in different blocks of the determinant (\ref{taublock}) \cite{YangYang3wave,Feng22b}. But that generalization is not necessary for our purpose.

Different ways to satisfy the dimension reduction condition (\ref{dimred}) will lead to different types of solutions to the Sasa-Satsuma equation. One type of such solutions --- rogue waves, were derived in \cite{Feng22b}. To derive rational solutions in Theorem~1, a different dimension reduction is needed. Following the $\mathcal{W}$-$p$ treatment we developed in \cite{YangYang3wave,YangYangBoussi}, we first introduce the function $\mathcal{Q}_{1}(p)$ as given in Eq.~(\ref{SasaQ1poly}), i.e.,
\[
\mathcal{Q}_{1}(p)=  \frac{1 }{p-{\rm{i}}\alpha} + \frac{1}{p+{\rm{i}}\alpha} +p,
\]
which is the coefficient of the exponential $\left(\partial_{r} + \partial_{s} + \partial_{x}\right)e^{\xi(p)}$. When $\alpha=1/2$ as in Theorem~1, the equation $\mathcal{Q}'_{1}(p)=0$ has a pair of double real roots $p_0=\pm \sqrt{3}/2$. In this case, we can show from \cite{YangYang3wave} that the dimension reduction condition (\ref{dimred}) would be satisfied if we choose
\[
\tau^{[I, J]}_{k,l}=
\left(
\phi_{3i-I, \, 3j-J}^{(k,l \hspace{0.04cm} I, J)}
\right)_{1\leq i \leq N_{I}, \, 1\leq j \leq N_{J}, \, p=p_0},
\]
\[
f_{1}(p) = \frac{\mathcal{W}_{1}(p)}{\mathcal{W}_{1}'(p)},
\]
the function $\mathcal{W}_{1}(p)$ is determined from the equation
\[ \label{3odeGeneSoluQ1}
\mathcal{Q}_{1}(p) = \frac{\mathcal{Q}_{1}(p_0)}{3} \left( \mathcal{W}_{1}(p) +\frac{2}{\sqrt{\mathcal{W}_{1}(p)}}
\cos\left[ \frac{\sqrt{3}}{2} \ln \mathcal{W}_{1}(p) \right] \right),
\]
and $f_2(q)=f_1(q)$. To introduce free parameters into these solutions, we choose $\xi_{0,I}(p)$ as
\[ \label{defxi0I2}
\xi_{0,I}=\sum _{r=1}^\infty \hat{a}_{r,I}  \ln^r \mathcal{W}_{1}(p), \quad I=1, 2,
\]
where $\hat{a}_{r,I}$ are free real constants.

Lastly, we simplify the matrix-element expression in Eq.~(\ref{Sasamijn}) and derive a more explicit expression without differential operators in it. This can be done by following the technique developed in \cite{OhtaJY2012,YangYang3wave}. Repeating such calculations, we then derive the solution formulae in Theorem~1, where free real parameters $a_{r,I}$ are related to $\hat{a}_{r,I}$ of (\ref{defxi0I2}) as
\[
a_{r,I}\equiv \hat{a}_{r,I}-b_{r},
\]
$b_r$ is the real expansion coefficient of the function
\[
\ln \left[ \frac{p \left(\kappa\right)+p_{0}}{2p_{0}} \right]= \sum_{r=1}^{\infty}b_{r} \kappa^r,
\]
and the real function $p(\kappa)$ is as defined in Eq.~(\ref{Q1ptriple}). It is noted that  Eq.~(\ref{Q1ptriple}) admits three branches of $p(\kappa)$ functions, which are related to each other as $p(\kappa e^{{\rm{i}}2j\pi/3})$, where $j=0, 1, 2$ (see Remark 3 in Ref.~\cite{YangYang3wave}). However, since $p(\kappa)$ in the current problem must be a real function, i.e., its Taylor expansion $p\left( \kappa \right)=\sum_{r=0}^{\infty}p_{r} \kappa^r$ must have real coefficients $p_r$, only one of those three branches is allowed.

\end{document}